 \documentclass[10pt,prd,aps,amssymb,  twocolumn,
 amsmath,tightenlines]{revtex4}
 \usepackage{graphics}
 \usepackage[pdftex]{graphicx}

 \usepackage{amsmath,amssymb,amsthm}
 \usepackage[utf8]{inputenc}
\newcommand{\snarkstanza}[1]{
    \begin{center}
    \begin{minipage}{0.87\textwidth}
        \begin{verse}
            \textit{#1}
        \end{verse}
    \end{minipage}
    \end{center}
    \vspace{0.7em}
}

\begin{document}
\title{ 
 Market Dynamics of Information Avalanches }
\author{B. K. Meister}
\date{\today}

\email{bernhard.k.meister@gmail.com}

\begin{abstract}
\noindent
Financial markets convert the incremental arrival of information into asset price changes. 
In a sandpile model grains of sand represent bits of data, and the size of an avalanche, governed by a scaling law,  is linked to price volatility.
While this model of self-organized criticality reproduces stylized facts, it also identifies a structural tension between the non-arbitrage condition and price adjustments consistent with a constant Sharpe ratio.
\end{abstract}
\maketitle
\section{Introduction}
\snarkstanza{
\hspace{-3.2em} He had bought a large map representing the sea,\\
\hspace{-3em} Without the least vestige of land:\\
\hspace{-3em} And the crew were much pleased when they found it to be\\
\hspace{-3em} A map they could all understand}

\noindent
A sandpile model\cite{bak} simulates the arrival and processing of  information. Bits of data replace grains of sand; otherwise the Bak-Tang-Wiesenfeld mechanism\footnote{ The Kuramoto model of synchronization offers an alternative to the Bak-Tang-Wiesenfeld sandpile. 
An early description of synchronization dynamics was given by Christiaan Huygens, who observed in 1665 that two pendulum clocks suspended from a wooden beam would soon move in `sympathy'. To test the coupling, he introduced obstructions and separated them  to desynchronize, only to find that, when brought back to the common support, the cadence re-established within half an hour.} stays unchanged.  Through a steady inflow  the sandpile is driven towards self-organised criticality (SOC). It leads to avalanches that follow a scaling law.   The step from informational propensities, to use a term coined by Popper, to market influence itself  is achieved via  avalanches, and 
temporarily bumps up price volatility. 
Therefore, a financial market  can be viewed as physical system that processes granular information. 

\noindent
The paper builds on the geometric foundations established in the companion paper\cite{meister2026b}, which showed that finite-resource observability forces markets into the exponential family. In the current paper, we look at Gaussians described in the one dimensional asset case by drift and volatility - a two dimensional manifold.
Here we ask: what are the dynamics on the manifold?

\noindent
 Desynchronization occurs, when market's rapid geodesic flow - propelled by information avalanches in a state of SOC -  meets the  lagging response of (central) bankers.
Markets can be viewed as multi-layered: 
\begin{description}
    \item[Layer 1: The Static View.] Static equilibrium -- a fixed point in the exponential family. 
    \item[Layer 2: The Sandpile Flow.] Motion on the equilibrium manifold. The system moves along paths  on the upper half-plane, with coordinates \((\mu,\sigma)\). Rapid traversals of these geodesics in the non-arbitrage case are responses to avalanches. The Bankers lag behind, 
    when volatility leaps. 
    \item[Layer 3: The Non-Equilibrium Extension.] Limited Onsager excursions slightly off the manifold, governed by detailed balance, which return via linear response.
\end{description}
Bankers' lag belongs to Layer 2. The market does not have to break down for conventional methods to fail.

\noindent
To simplify, we introduce a slowly varying Sharpe ratio that ties excess return to volatility. A doubling of volatility doubles the excess return, which -- assuming a slowly adjusting long-run target price -- implies a sudden price drop when volatility spikes. Time scales matter: avalanches happen suddenly; the Sharpe ratio and long-run target price adjust relatively slowly. This provides a rationale why downward corrections are often sudden.

\noindent
The mapping of avalanche size to  the volatility $\sigma$ through averaging, sliding windows or weight functions provides many choices,  and in particular influences option valuations. We forego analysing derivatives and instead look at `arbitrage'. This arises from the different path the drift and volatility can take on the manifold between relatively stable configurations.
The point on the manifold normally creeps along, but is occasionally jolted  by a burst of incoming information leading to a leap in volatility. How does the market state move from the initial point in $\sigma$ and $\mu$ space to this new point? It can follow a geodesic path, which is selected out, because it has an extremal property, or deviate from it.   
It will be shown that deviations from the geodesic, e.g. by following a Sharpe ratio constant path, provide profit opportunities. 

\noindent
Each section is introduced by a stanza from Lewis Carroll's \emph{The Hunting of the Snark}. 

\noindent
The paper is organized as follows. Section II develops the sandpile model. Section III introduces the geometry of the upper half-plane and the geodesic equations. Section IV states the geodesic no-arbitrage relation. Section V quantifies the arbitrage potential between geodesic and Euclidean paths and derives the optimal dynamic trading strategy. Section VI distinguishes geodesic flow from Onsager excursions.   Section VII concludes. 

\section{Sandpile Model}
\snarkstanza{
   \hspace{-0.6em} They sought it with thimbles, they sought it with care;\\
   They pursued it with forks and hope;\\
They threatened its life with a railway-share;\\
   They charmed it with smiles and soap.}
   
\noindent
Information is modeled as a sandpile in a state of self-organized criticality.
Each discrete event -- a trade, a news announcement, a data release -- adds a grain to an accumulating pile. The pile's slope measures unrevealed information, waiting to be integrated in the price. When the slope reaches a critical threshold, the chance of an additional grain  to trigger an avalanche is high. Information floods the market. 

\noindent
Two features of importance:
     \textbf{Information loading} happens slowly and deterministically; i.e. grain by grain.
     \textbf{Avalanches} happen suddenly and reset the information gradient.

\noindent
Incremental data events   accumulate to form an information gradient, the system's height of unprocessed information, and corresponds to the pile's angle \(\theta(t)\).
Between avalanches, the information gradient builds at rate $v$:
\begin{equation}
    d\theta = v dt \label{eq:accumulation}\nonumber
\end{equation}
As the critical threshold \(\theta_c\) is approached, the probability for an avalanche of size \(s\) to occur, resetting \(\theta\) to a lower value, increases.  The probability for avalanche size   \(s\) is proportional to the overshoot:
\begin{equation}
    s \propto \alpha(\theta - \theta_c). \label{eq:size}\nonumber
\end{equation}
After the avalanche, the cycle repeats. This is a mean-reverting process with jumps\footnote{The relationship between avalanche size $s$ and volatility is captured by the slope dynamics: $d\theta_t = v dt - s(\theta_t) dN(\theta_t)$, where $v$ is the steady information inflow. As $\theta_t \to \theta_c$, the state-dependent coupling of the jump probability $dN(\theta_t)$ and the magnitude $s(\theta_t)$ ensures the system exhibits the characteristic scaling laws of SOC, pinning the dynamics to the critical threshold.
}. 

\noindent
In the financial context, the avalanche size \(s\) drives volatility. When an avalanche occurs, volatility jumps; between avalanches, it mean-reverts. This asymmetry--volatility jumps up instantaneously and decays slowly--is the  signature of information avalanches. It distinguishes the model from symmetric jump processes. 
The exact mapping of avalanche size $s$ to  the volatility $\sigma$ through averaging, sliding windows or a weight function we forego, since any `reasonable' mapping\footnote{
By reasonable' we mean any mapping that is monotonic in $s$, does not use future information, and allows  avalanches to bump volatility discontinuously, but 
the exact functional form does not affect the existence of the geodesic vs. Euclidean path difference, which is the focus of this paper. The discontinuity is essential, because it forces the interpolation between two distinct states.}
will lead to the same result. 
What matters is the qualitative picture: stress builds, stress releases, and releases are fast.
\noindent

\section{The Geometry of the Upper Half-Plane}
\snarkstanza{The Butcher rendered ideas of his own,\\
With a somewhat peculiar grace:\\
And instructed the Snark on the path it should take,\\
By the light of the sun and the face.}


\noindent
The statistical manifold for a normal distribution is the upper half-plane  with coordinates.
\begin{equation}
    x^1 = \mu \quad (\text{drift}), \qquad x^2 = \sigma \quad (\text{volatility}).\nonumber
\end{equation}
The Fisher information metric (up to a global scaling factor\footnote{   Note, the parameters $(\mu, \sigma)$, where  drift $\mu$ normally scales as $T^{-1}$ and volatility $\sigma$ as $T^{-1/2}$,  are de-dimensionalized by the characteristic information-arrival window to resolve any temporal scaling discrepancies.})   is
\begin{equation}
    ds^2 = \frac{d\mu^2 + 2d\sigma^2}{\sigma^2}.\nonumber
\end{equation}
The metric is non-flat, possessing a constant negative curvature; this renders Euclidean straight-line paths non-geodesic, a fact of importance in subsequent sections.
The Christoffel symbols  are
\begin{equation}
    \Gamma^\mu_{\mu\sigma} = -\frac{1}{\sigma},\quad  \nonumber
    \Gamma^\sigma_{\mu\mu} = \frac{1}{2\sigma},\quad \nonumber
    \Gamma^\sigma_{\sigma\sigma} = -\frac{1}{\sigma}. \label{eq:christoffel}\nonumber
\end{equation}
For Gaussians, the Fisher information metric, the Christoffel symbols, and the resulting semicircular geodesic equations,  confirming its structure as the hyperbolic Poincare upper half-plane, as well as the path length of importance in the next two sections  can all be found in Amari's book\cite{amari2016}.

\noindent

\noindent

\noindent

\noindent

\noindent
Imagine an avalanche occurs. This leads  not to a departure from the equilibrium manifold, but it is an extremely rapid traversal of a particular path. The system moves from \((\mu_1,\sigma_1)\) to \((\mu_2,\sigma_2)\) along a geodesic, i.e. a minimal length path. 
\noindent
The geodesic connecting two points is given by the semicircle
\begin{equation}
    (\mu-\mu_c)^2 + 2\sigma^2 = R^2, \label{eq:geodesic} 
\end{equation}
where the center \(\mu_c\) and radius \(R\) are  determined by the  endpoints:
\begin{align}
    \mu_c &= \frac{\mu_1+\mu_2}{2} + \frac{\sigma_1^2-\sigma_2^2}{\mu_1-\mu_2},\nonumber \\
    R &= \sqrt{(\mu_1-\mu_c)^2 + 2\sigma_1^2}.\nonumber
\end{align}
The resulting path is the information-geometric analogue of a brachistochrone\footnote{
Just as Bernoulli derived the global Brachistochrone by integrating the local refraction (Snell’s law) into Fermat’s Principle of Least Time, the Fisher-Rao geodesic integrates the infinitesimal   Kullback-Leibler divergence (the Fisher information metric) into a global minimal distance.}.


\noindent

\noindent
The Banker's lag occurs in Layer 2, because the market moves rapidly within the  equilibrium manifold. The geometry still holds; the coordinates still transform via geodesics; the exponential family still describes the state, but a central banker's responses normally would lag. Crashes are therefore not necessarily breakdowns of equilibrium. Instead, they  can be due to rapid traversals of the equilibrium manifold\footnote{The implied volatility smile emerges as a projection of the geodesic onto option strike space. 
This connects to the maximum entropy option pricing of \cite{Brody}.}.

 
\noindent
 

 \section{No-Arbitrage Geodesics}
 \snarkstanza{
  \hspace{-0.6em}   And the Banker, inspired with a courage so new --\\
    It was matter for general remark,\\
    Rushed madly ahead and was lost to their view\\
    In his zeal to discover the Snark
}

 \noindent

   \noindent
The market evolves from an initial to a final  market state  as a response to an avalanche. This update must satisfy the principle of Minimum Information Action\footnote{
The principle of `Minimum Information Action' is the information-theoretic equivalent  of the principle of least action in classical mechanics. 
It dictates that the system moves between equilibria $(\mu_1, \sigma_1)$ and $(\mu_2, \sigma_2)$ along the geodesic -- the unique path of zero dissipation ($dQ=0$).  Suboptimal path-selection generates `information heat', and any deviation from geodesic implies that the market is `performing' unnecessary informational work and results in potentially harvestable arbitrage -- see \cite{meister2026b} 
} and selects out a geodesic, which represents a  reversible, zero-dissipation evolution.  Any other path generates `information heat' ($dQ > 0$)  and an arbitrage potential.

\noindent
\noindent

\noindent


\noindent
The no-arbitrage condition forces the response to the avalanche to follow the unique geodesic -- the path of zero dissipation. The geodesic length is the hyperbolic distance:
\begin{equation}
    L_{\text{geo}} = \operatorname{arcosh}\left(1 + \frac{(\mu_2-\mu_1)^2 + 2(\sigma_2-\sigma_1)^2}{4\sigma_1\sigma_2}\right).
    \label{eq:geo_length}
\end{equation}

\noindent
The existence of avalanches also reprices options. In the presence of Self-Organized Criticality (SOC), avalanche size $s$ scales as $s^{-\tau}$ and creates heavy-tailed option pricing\footnote{Option valuation is given by $\int \frac{\sigma^2}{2} \Gamma \, dt$. In the extreme case of  roughly $\tau < 2$ and a `reasonable' mapping into $\sigma$ the $s^{-\tau}$ scaling of the SOC avalanche size  leads to an unlimited volatility-driven gamma rent, transforming `informational heat' into a wealth divergence. The paper  forgoes a comparison of SOC option pricing with rough volatility, wandering gaussians or other models to avoid having to justify how to map avalanche size to $\sigma$.\\
One must distinguish between two types of `information heat', i.e.  `arbitrage heat' and `geometric heat'. Arbitrage heat is dynamic and path-dependent; it is the `wealth leak' from choosing a linear ray instead of a geodesic. Geometric heat is the inescapable cost of the information flow itself, and determines option values.  Restoring correct time dependence clarifies sensitivity to time. Arbitrage Heat scales linearly, while the Geometric Heat scales as the square root of time.}.

\noindent
 The Sharpe ratio\footnote{Only one asset is directly considered, but this asset is assumed to be part of a wider market. For simplicity the risk-free rate is set to zero, $r=0$.} 
 \begin{equation}
    S = \frac{\mu_1}{\sigma_1},\nonumber
\end{equation}
 is a natural market invariant that allows assets and portfolios to be compared. For a transition between states sharing the same Sharpe ratio, $S$ defines both the coordinate ray from the origin and the slope of the linear chord between them.  
 However, the Sharpe ratio conserving dynamics do not coincide with the geodesic. 
 As an example, consider the `linear' path with constant Sharpe ratios that  switches from \((\mu_1,\sigma_1)\) to \((\mu_2,\sigma_2)\), 
  and the trajectory is
\begin{equation}
    \sigma(f) = \sigma_1 + f(\sigma_2-\sigma_1),\quad \mu(f) = \sigma(f)S + r,\quad f\in[0,1].\nonumber
\end{equation}
Its length in the Fisher metric is
\begin{equation}
    L_\text{lin} = \int_0^1 \frac{\sqrt{(\frac{d\mu}{df})^2 + 2(\frac{d\sigma}{df})^2}}{\sigma(f)} df = \sqrt{S^2+2}\ln\frac{\sigma_2}{\sigma_1}.
    \label{eq:lin_length}\nonumber
\end{equation}
 The discrepancy   $\Delta L_t = L_{\text{lin}} - L_{\text{geo}}$  is the source of the dynamic arbitrage strategy,  $  \int W_t \, \Delta L_t \, dt$, where $W_t$ is the capital employed, and evaluated in a discrete approximation in the beginning of the following section.

  \noindent
While the existence of $\Delta L > 0$ identifies and quantifies an opportunity, is it is not a sufficient condition for risk-less profit.
 In real markets, frictions and constraints may   prevent exploitation. Nevertheless, the geodesic is the unique path consistent `non-arbitrage'.
 The actual exploitation also depends on the arbitrageurs  ability to navigate the manifold faster than the dissipation of the information signal. Otherwise, the heat might simply dissipate as noise.


    \noindent


 \noindent

\noindent

\noindent


\noindent
 \section{The Dynamic Arbitrage Strategy}
\snarkstanza{
 \hspace{-0.6em} The Bellman looked frightened, and turned very pale,\\
And let the bell drop from his hand:\\
And his voice trembled so, that he could hardly say\\
The first thing he’d planned to say.}

\noindent
The geodesic no-arbitrage relation establishes that any transition between equilibrium states must follow the Fisher-Rao geodesic to avoid `information heat'. But this heat is not merely a theoretical abstraction—it represents a profit opportunity. We now quantify the excess action of the Euclidean (constant Sharpe) path and derive the optimal dynamic strategy to capture it.

\noindent
Consider an infinitesimal step from \((\mu,\sigma)\) to \((\mu+d\mu,\sigma+d\sigma)\) with \(d\sigma = \varepsilon\) small and \(d\mu = S\varepsilon\) following the constant Sharpe constraint. The linear path length in the Fisher metric is
\begin{equation}
L_{\text{lin}} = \sqrt{S^2+2}\,\frac{\epsilon}{\sigma} + \mathcal{O}(\varepsilon^2)\nonumber. 
\end{equation}

\noindent
The geodesic connecting these nearby points is given by the hyperbolic distance. 
The expansion of the `arcosh' function,
\begin{equation}
\operatorname{arcosh}(1+x) =
 \sqrt{2x} + \mathcal{O}(x^{3/2}), \nonumber
 \end{equation}
 with 
 \begin{equation}
x = \frac{(S^2+2)\varepsilon^2}{4\sigma(\sigma+\varepsilon)} = \frac{(S^2+2)\varepsilon^2}{4\sigma^2}+ \mathcal{O}(\epsilon^{3 }),\nonumber
\end{equation}
gives
\begin{equation}
L_{\text{geo}} = \sqrt{\frac{S^2+2}{2}}\,\frac{\varepsilon}{\sigma} + \mathcal{O}(\varepsilon^2).  \nonumber
\end{equation}

\noindent
The excess action for this infinitesimal step is therefore
\begin{equation}
\Delta L = L_{\text{lin}} - L_{\text{geo}} = \sqrt{S^2+2}\left(1 - \frac{1}{\sqrt{2}}\right)\frac{\varepsilon}{\sigma} + \mathcal{O}(\varepsilon^2),  \nonumber
\end{equation}
and represents the arbitrage potential. 
 In an ideal market it would be eliminated by forcing the geodesic. But if the market temporarily follows the Euclidean path,  \(\Delta L\) becomes a potentially harvestable profit.
Even if the arbitrageur does not   know the endpoint \((\mu_2,\sigma_2)\), the opportunity is exploitable, since the strategy is    local and dynamic.


\noindent
At any instant, the current market state \((\mu,\sigma)\) defines a local geodesic direction via the geodesic equation:
\begin{equation}
\frac{d\mu}{d\sigma} = -\frac{2\sigma}{\mu-\mu_c},\nonumber
\end{equation}
where \(\mu_c\) is the (unknown) center of the full geodesic. The local slope can be inferred from the current curvature of the manifold.

\noindent
The mispricing between the Euclidean expectation (constant Sharpe) and the  geodesic creates instantaneous hedge ratios. For any derivative \(V(\mu,\sigma,t)\), the difference in its predicted change along the two paths is
\begin{equation}
\delta V = \frac{\partial V}{\partial \mu}(\mu_{\text{geo}}-\mu_{\text{lin}}) + \frac{\partial V}{\partial \sigma}(\sigma_{\text{geo}}-\sigma_{\text{lin}}).\nonumber
\end{equation}

\noindent
To be neutral to the geodesic, i.e. to have zero exposure if the market follows the geodesic, while being exposed to the Euclidean path, the arbitrageur holds:
\begin{itemize}
    \item Long the underlying  in proportion to \(\partial V/\partial \mu\),
    \item  Short volatility  (Vega) in proportion to \(\partial V/\partial \sigma\).
\end{itemize}
The ratio of Vega short to stock long at any instant is
\begin{equation}
\frac{\text{Vega position}}{\text{Stock position}} = \frac{\partial V/\partial \sigma}{\partial V/\partial \mu} = \frac{\nu}{\Delta}, \nonumber
\end{equation}
where \(\nu\) is Vega and \(\Delta\) is delta. This ratio evolves along the path according to the geodesic curvature.
Re-balancing continuously as the market updates its state, the arbitrageur harvests  \(\Delta L\) increment by increment, 
 and is analogous to running a  Carnot engine (for an application of Carnot cycles to finance see \cite{Meister2026a})  between the initial and final states. The  geodesic  is the reversible, zero-entropy path. The  Euclidean path  corresponds to an irreversible, heat-generating process. The arbitrageur's aim is to extract as much as possible of the excess heat.
Heat is not a modeling artifact; it is the fundamental cost of moving through information space in a suboptimal way. 

\noindent

\noindent

\section{Geodesic Flow and Onsager Excursions}
\snarkstanza{
   \hspace{-2.9em} Other maps are such shapes, with their islands and capes!\\
  \hspace{-2.6em} But we've got our brave Captain to thank\\
 \hspace{-2.6em} (So the crew would protest) `that he's bought us the best-\\
 \hspace{-2.6em} A perfect and absolute blank!'\\
}


\noindent
The companion paper established that finite-resource markets are described by a  minimal exponential family -- an equilibrium geometry.  This section outlines two different ways of market movement.

\noindent
{\bf  Layer 2}: Geodesic Flow on the Equilibrium Manifold

\noindent
The upper half-plane with coordinates \((\mu,\sigma)\) carries the Fisher metric. 
The avalanche is an extremely rapid traversal of a geodesic segment. 
The system never leaves the equilibrium manifold. 

\noindent
{\bf Layer 3}: Onsager Excursions (Limited Departures)

\noindent
When the system is pushed slightly off the equilibrium manifold -- by a perturbation that violates the geodesic flow -- it enters the linear response regime governed by Onsager reciprocity (see \cite{meister2026b} for an application to finance):
\begin{equation}
     \dot{\eta}_i  = \sum_j L_{ij} X^j, \nonumber
\end{equation}
where \(X^j  \) are thermodynamic forces and \(L_{ij}\) is the transport matrix. 
This is a limited, controlled departure from equilibrium -- a transient perturbation. 


\section{Conclusion: Banker's Fate} 
\vspace{.18cm}
\snarkstanza{He offered large discount -- he offered a cheque\\
(Drawn to bearer') for seven-pounds-ten:\\
But the Bandersnatch merely extended its neck\\
And grabbed at the Banker again.\\
\vspace{.8cm}}

 \vspace{.8cm}

\noindent
 \noindent
 Banker’s lag is the consequence of minimizing the time spent violating the Sharpe-constancy macro-constraint. By hurtling along the zero-dissipation geodesic to eliminate micro-arbitrage during an informational avalanche, the market outruns the Banker’s capacity to respond and results in a state of market desynchronization.
This is not a failure of the
equilibrium. The minimal exponential families are rich enough to host both static equilibrium and market corrections, proving there is no necessity to invoke non-exponential families to study extreme market events. 


\noindent
The dynamic arbitrage strategy derived from this geometry provides a possible way to harvest the  
excess  when markets deviate from the geodesic, and the three-layer framework unifies the static equilibrium of the companion paper\cite{meister2026b} with the dynamics of stress accumulation and release.

\noindent

\noindent
Many questions remain: empirical calibration of the stress gap,   the  functional form of the loading rate $v$ and the jump process, and more generally the thermodynamic signature of the information avalanches. 

\noindent

\noindent
The Bandersnatch  is already contained in the  minimal map  of exponential families navigated by the banker. There is no requirement to extend the map to understand the territory. The elusive Snark of the equilibrium can already force the Banker's Lag. 

\end{document}